# A Digital Watermarking Approach Based on DCT Domain Combining QR Code and Chaotic Theory

Qingbo Kang[(1)], Ke Li[(2)], Jichun Yang[(2)]
1 Chengdu Yufei Information Engineering Co.,Ltd., Chengdu, China
2 National Key Laboratory of Fundamental Science on Synthetic Vision, Sichuan University, Chengdu, China
E-mail: qdsclove@gmail.com

*Abstract*—**This paper proposes a robust watermarking approach based on Discrete Cosine Transform (DCT) domain that combines Quick Response (QR) Code and chaotic system. When embed the watermark, the high error correction performance and the strong decoding capability of QR Code are utilized to decode the text watermark information which improves the robustness of the watermarking algorithm. Then the QR Code image is encrypted with chaotic system to enhance the security of this approach. Finally the encrypted image is embedded to the carrier image's DCT blocks after they underwent block-based Arnold scrambling transformation. During the extraction process, as long as the QR Code image can be decoded, the completeness and accuracy of the text watermarking information can be guaranteed. The results of simulation experiment show that this approach has high robustness and security and has, therefore, some practical value in the copyright protection.**

*Keywords—QR Code; digital watermarking; information hiding; copyright protection*

## I. INTRODUCTION

With the rapid development of computer networks and the Internet, the high-speed transmission, processing and storage of digital media had become a reality. However, these advances have also made it possible to copy and modify it. Therefore the copyright protection for digital media has received widespread attentions [1]. Digital watermarking technique, as a kind of new technique of copyright protection and content authentication for digital media, has been extensively researched, and many watermarking schemes have been proposed [2-4]. In general, a digital watermark is some information about the digital media's copyright owner and intended to be permanently embedded into the digital media [5]. According to the different types of protected digital media, digital watermarking can be divided into image watermarking, audio watermarking, video watermarking, etc. This paper focuses on image watermarking, which has two conflicting requirements: imperceptibility and robustness. The conflicting point is how to improve the watermark's ability against the attacks while influencing as less as possible the original image. Hsieh proposed a digital watermarking algorithm based on Discrete Cosine Transform (DCT) domain, the robustness of which is high to most attacks, but is low to JPEG compression [6].

This paper presents a novel approach based on DCT domain that combines Quick Response (QR) Code technique. Due to the high error correction performance and strong decoding capability of QR Code, when it is adopted as the watermark information, this approach is given a certain degree of error correction capability, so that the robustness of watermarking could be improved. Then the original QR Code image is encrypted with chaotic system to reinforce the security. Furthermore, this approach is a blind watermarking technique, which means that the extraction process of it does not require the original image. The following sections describe the watermarking approach in detail, the simulation experiment shows that this watermarking approach is safe, reliable, and has a good invisibility as well as strong robustness for some general image processing methods such as lossy compression, noise pollution, image filtering, image modification and image enhancement.

## II. PRETREATMENT OF WATERMARK INFORMATION

### A. Generating chaotic sequence

Chaotic sequence is of highly unpredictable and random-look nature. It is generated by chaotic systems, which have the following three properties [7]:

*1) Deterministic, this means they have some determining mathematical equations ruling their behavior.*

*2) Unpredictable and non-linear, this means they are sensitive to initial conditions. Even a very slight change in the starting point can lead to significant different outcomes.*

*3) Appear to be random and disorderly but in actal fact they are not. Beneath the random behavior there is a sense of order pattern.*

To increase the security and the robustness of the watermark image, chaotic sequence is used to encrypt it, so as to decrease its spatial correlation.

Logistic map is one of the simplest and most transparent systems exhibiting order to chaos transition. Mathematically it is defined as:

$$x_{n+1} = \mu x_n(1-x_n), n \in Z, \mu \in [0,4], x_n \in (0,1) \quad (1)$$

The $\mu$ here is a positive constant sometimes known as the "biotic potential", when $3.5699456 < \mu < 4$ the map is in the region of fully developed chaos [8]. That is, at this point, the sequence $\{x_k; k=0,1,2,3,...\}$ generated by (1) is non-periodic, non-convergent and sensitive to the initial value. So using the sequence to encrypt the watermark signal can give the watermark system a very strong anti-decipher capacity.

*B. Encoding watermark information with QR Code*

QR Code is a two-dimensional bar code in the form of the Matrix Code that invented in 1994 by Denso Wave. It has greater storage capacity, higher density, stronger error correction performance and safety over one-dimensional bar code. Information that can be stored in QR Code is not only text information such as numbers and characters, but also the high-capacity information like images [9].

Considering the above advantages of QR Code, taking QR Code image as the watermark image will give the watermarking system a certain degree of error correction ability and improve the robustness. Therefore in the proposed approach, the text watermark information is firstly encoded with QR Code encoding algorithm to generate the QR Code image, which is then embedded into the carrier image.

*C. Encrypting the QR Code image with chaotic sequence*

After generating the text watermark information's QR Code image, in order to guarantee the safety, decrease the spatial correlation of the embedded watermark image and avoid of blocking artifacts (a distortion that appears in compressed image material as abnormally large pixel blocks) [10], the logistic map is used to generated a chaotic sequence to encrypt the QR Code image.

Since the QR Code images are always squares and two-valued, let us denote the original text watermark information's QR Code image as $Q$, the size of which is $N \times N$, hence $Q(i,j) = \{0,1\}$, where $i \in [0, N-1]$, $j \in [0, N-1]$. The encryption process is described as below:

*1) Giving the initial value $x_0$ and the constant value $\mu$, the logistic map defined in the formula (1) generates the logistic sequence $X$ the length of which is $N \times N$:*

$$X = \{x_0, x_1, x_2, ..., x_{N \times N-1}\}$$

*2) The logistic sequence generates the binary sequence using the threshold value division method (the threshold value used here is 0.5) [11]:*

$$b_k = \begin{cases} 0, x_k < 0.5 \\ 1, x_k \geq 0.5 \end{cases}, k \in [0,1,2,...,N \times N-1]$$

So the sequence $B = \{b_0, b_1, b_2, ..., b_{N \times N-1}\}$ is binarized.

*3) Performing XOR operation with the binary sequence $B$ and the binary image $X$ to generate the encrypted sequence $E$:*

$$E = \{e_0, e_1, e_2, ..., e_{N \times N-1}\}$$

After these above steps, the security of watermark information is enhanced. Fig. 1(a), displays the original QR Code image (the content of QR Code is "Watermark information"). And Fig. 1(b), shows the encrypted image (the initial value is 0.6 and the $\mu$ is 3.99).

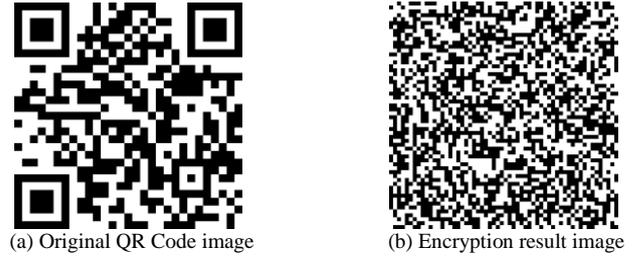

(a) Original QR Code image      (b) Encryption result image

Fig. 1. Original QR Code and the encryption result

During the extraction process of watermark information, the initial value and logistic parameter $\mu$ used in Step 1 will be needed to recover the original QR Code image, which is decoded to obtain the text content in QR Code. So only the authorized person who knows these values can be permitted to get the text watermark information. This method ensures the security of this watermarking approach.

### III. THE PROPOSED APPROACH

Currently the digital watermarking approaches can be divided into two categories: the spatial domain based and the transform domain based. Approaches based on transform domain hide the watermarking data in transform coefficients, therefore spreading the data through the frequency spectrum, making it hard to detect and strong against many types of signal processing manipulations [12]. At present, most robust watermarking approaches are based on transform domain.

DCT is one of the most commonly used linear transform in digital signal processing. Also it's used in lossy compression of image (e.g. JPEG). The watermarking approach based on DCT domain can enhance the ability to resist JPEG lossy compression. Therefore, researches in this respect have received a widespread attention [6]. Our approach also embeds the watermark information in DCT domain.

*A. Watermark Embedding Process*

The watermark embedding process is show in Fig. 2.

Assuming the original carrier image is $I$, and its size is $M \times M$, the encrypted watermark image is W, which size is $N \times N$. The specific process of watermark embedding describes as below:

**Step 1:** As well as the JPEG standard, we split the whole original carrier image into blocks of $8 \times 8$ pixels. So we have a total of $\dfrac{M \times M}{64}$ blocks. Wherein the block which the position is $(i, j)$ in spatial domain is $S_{ij}(x, y)$, where $x, y \in \{0,1,...,7\}$ and $i, j \in \{0,1,..., \dfrac{M}{8}-1\}$.

**Step 2:** Considering the attack of regional modification for watermarked image and the security of our approach, next scramble these blocks with Arnold transform, i.e. Block-based Arnold scrambling is applied to the original image.

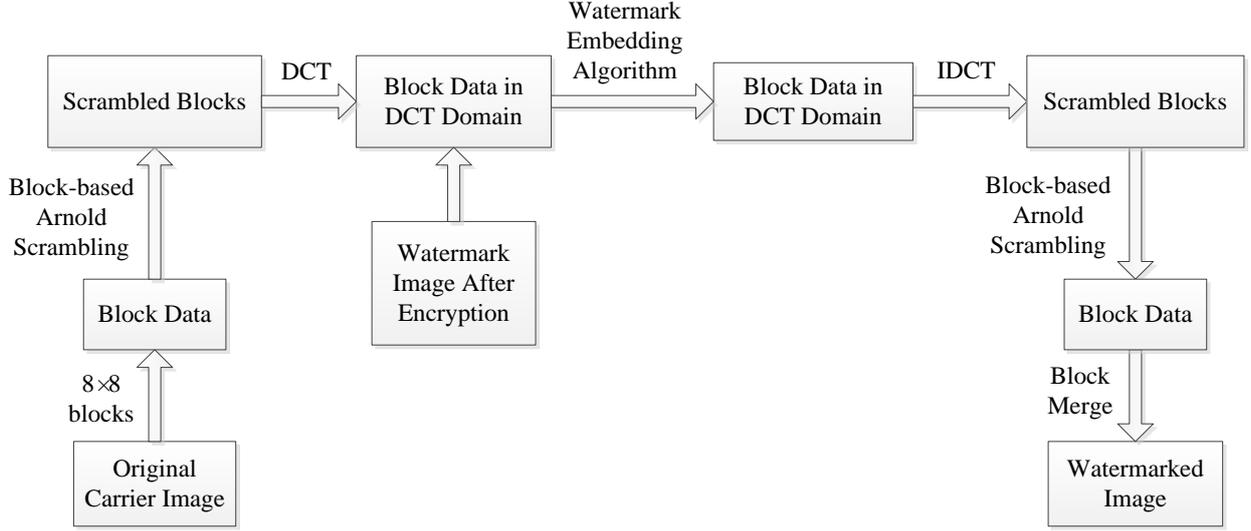

Fig. 2. Watermark embedding process.

Arnold transform is proposed by V.I. Arnold in the research of ergodic theory [13]. It is also a kind of widely used geometry transform based on pixel scrambling invertible method. We use the two-dimension form of it as our image block-based scramble method, the formula is show in Eq. (2).

$$\begin{bmatrix} i' \\ j' \end{bmatrix} = \begin{bmatrix} 1 & 1 \\ 1 & 2 \end{bmatrix} \begin{bmatrix} i \\ j \end{bmatrix} (\mod(\frac{M}{8})) \quad i, j \in \{0, 1, ..., \frac{M}{8} - 1\} \quad (2)$$

Where $i$ and $j$ are the block position of original image, $i'$ and $j'$ are the block position of scrambled image. Arnold transform is a periodicity transform, the period is depend on the size, i.e. $\frac{M}{8}$ in this case.

**Step 3:** Transform each block in frequency domain using DCT and apply zig-zag scan to all 64 DCT coefficients. Denote the number $k$ block's in DCT domain as:

$$F_k(u, v) = zig\_zag(DCT\{S_k(x, y)\})$$

Where $u, v, x, y \in \{0, 1, 2, ..., 7\}$, and $k$ is in the range $[0, 1, 2, ..., \frac{M \times M}{64})$.

**Step 4:** Determining the embedding positions in DCT domain. In order to keep a balance of the imperceptibility and the robustness of watermarking, our approach choose the middle band frequency coefficients to embedded watermark, the reasons are: (1) The low band frequency concentrated the most energy of an image, embed in this region will seriously affect the visual quality of the image, the human eye can perceive changes in the image, doesn't meet the requirement of imperceptibility; (2) The high band frequency is the most easily removed region to plenty of image processing methods such as low-pass filter, lossy compression and image noise, doesn't satisfied the requirement of robustness [5].

The middle band frequency coefficients in one block numbered with zig-zag scan order shows in Fig. 3.

|   |   |   |   | 1 | 9 | 10 | 22 |
|---|---|---|---|---|---|----|----|
|   |   |   | 2 | 8 | 11 | 21 |   |
|   |   | 3 | 7 | 12 | 20 |   |   |
|   | 4 | 6 | 13 | 19 |   |   |   |
|   | 5 | 14 | 18 |   |   |   |   |
| 15 | 17 |   |   |   |   |   |   |
| 16 |   |   |   |   |   |   |   |
|   |   |   |   |   |   |   |   |

Fig. 3. Middle band frequency coefficients in DCT domain.

The shadow region in Fig. 3, is the middle band frequency. Next, choose three continuous coefficients in this region and ten coefficients adjacent to them. Based on the Human Visual System (HVS) model proposed in [14], embedding the watermark signals in the range (6, 14) will not cause too much change to human eyes. Denote the three coefficients chosen in the number $k$ block are $x_k(i-1), x_k(i), x_k(i+1)$, so the ten adjacent coefficients are $\{x_k(j), j \in \{i-6, i-5, i-4, i-3, i-2, i+2, i+3, i+4, i+5, i+6\}\}$.

**Step 5:** Watermark embedding. In our approach, we embed the watermark by using the coefficients modification method. Specifically describes below:

(1) Calculate the average value of the ten coefficients, that is:

$$\overline{x_k} = \frac{1}{10} \sum_j (x_k(j), \ j \in \{i-6, i-5, i-4, i-3,$$
$$i-2, i+2, i+3, i+4, i+5, i+6\})$$

(2) Embed watermark. In our approach, we embed one bit watermark information in one $8 \times 8$ block. Due to QR Code images are binary images, as well as the encrypted image of original QR Code image is always binary, let $W$ indicate the encrypted watermark image, so $W = \{w_k, w_k \in \{0,1\}\}$. The embedding positions are $x_k(i-1)$, $x_k(i)$, $x_k(i+1)$, and the formula is:

$$\begin{cases} x_k(i-1) = x_k(i) = x_k(i+1) = \overline{x_k} + \lambda, w_k = 1 \\ x_k(i-1) = x_k(i) = x_k(i+1) = \overline{x_k} - \lambda, w_k = 0 \end{cases}$$

Where $w_k$ is the bit in the corresponding position of the encrypted watermark image, $\lambda$ is the strength for watermark embedding. The bigger the $\lambda$ is, the much robust of the watermark ability against to the attacks, but the more poor of the imperceptibility for carrier image.

**Step 6:** Applying Inverse Discrete Cosine Transform (IDCT) to each block to obtain the block data contains watermark information in spatial domain.

**Step 7:** Block based Arnold scrambling described in Eq. (2).

**Step 8:** Merging these blocks in spatial domain, then we can obtain the watermarked image $I^*$ which size is $M \times M$.

*B. Watermark Extracting Process*

The proposed watermarking approach is a kind of blind watermarking approach, which means the watermark extracting doesn't need the original image [6]. The process of extraction is describes below:

**Step 1:** Splitting the watermarked image into $8 \times 8$ blocks $P_{ij}(x, y)$, where $x, y \in [0, 7]$ and $i, j \in [0, \frac{M}{8})$.

**Step 2:** Block based Arnold scrambling described in Eq. (2).

**Step 3:** Applying DCT to each image block $P_{i'j'}$.

**Step 4:** Locating the three middle band coefficients in embedding process. In block $k$, we denote these values by $x_k^*(i-1)$, $x_k^*(i)$, $x_k^*(i+1)$. And calculating the average of the ten coefficients adjacent to the three values, denote the average by $\overline{x_k^*}$.

**Step 5:** Watermark extracting. The number of coefficients in the three $x_k^*(i-1)$, $x_k^*(i)$, $x_k^*(i+1)$ is greater than or equal to $\overline{x_k^*}$ is denoted by $count\_max$, and less than is denoted by $count\_min$, $w_k'$ is the extract watermarking bit in block $k$, so:

if $count\_max \geq count\_min$, then $w_k^* = 1$;

if $count\_max < count\_min$, then $w_k^* = 0$.

Applying this operating on all blocks, then we can extract the encrypted watermark image $W^*$.

**Step 6:** Decrypting $W^*$ then we can obtain the original QR Code image. Then using QR Code decoding algorithm to decode the text watermark information..

IV. SIMULATION EXPERIMENTS

*A. Evaluation criterias*

Generally speaking, evaluating a watermarking approach mainly includes two parts: robustness and imperceptibility. The robustness is objective judged by Normalized Correlation (NC) value between the original watermark $W$ and the extracted watermark $W^*$. NC is defined as Eq. (3).

$$NC = \frac{\sum_{i=1}^{M_1} \sum_{j=1}^{M_2} W(i,j) W^*(i,j)}{\sqrt{\sum_{i=1}^{M_1} \sum_{j=1}^{M_2} [W(i,j)]^2} \sqrt{\sum_{i=1}^{M_1} \sum_{j=1}^{M_2} [W^*(i,j)]^2}} \quad (3)$$

The NC value is between 0 and 1, the more closely to 1, the better robustness of the watermarking approach [15]. In addition, because the watermark in our approach is QR Code image, so whether the extracted QR Code can be decoded is another important judgment criterion.

The visual quality of watermarked image (i.e. imperceptibility) is evaluated by the peak signal-to-noise ratio (PSNR) criterion defined as Eq. (4).

$$PSNR = 10 \log_{10} \frac{A^2}{\frac{1}{N \times M} \sum_{i=1}^{N} \sum_{j=1}^{M} [f(i,j) - f^*(i,j)]^2} \quad (4)$$

In Eq. (4), $f(i,j)$ is the original image pixel value at coordinate $(i,j)$ and $f^*(i,j)$ is the altered image pixel value at coordinate $(i,j)$. $A$ is the largest energy of the image pixels (i.e., $A = 255$ for 256 gray-level images) [16]. The unit of PSNR is dB. And the bigger the PSNR value, the better imperceptibility of the watermarking approach.

*B. Simulation results*

The carrier image selected in simulation experiments is a stand 512×512 grayscale bitmap Lena, watermark information is the text "Watermark information". Fig. 5(a), shows the original carrier image, the image after block based Arnold scrambling is shown in Fig. 5(b), the times of Arnold scrambling is 20.

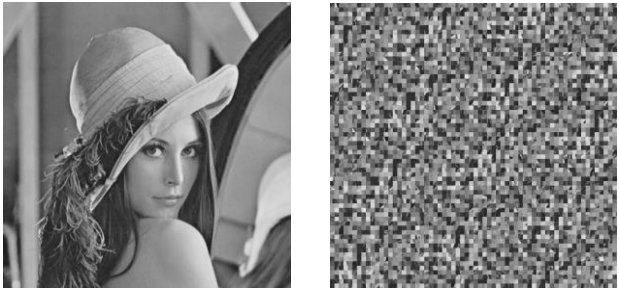

(a) Original carrier image Lena    (b) The scrambled image of Lena

Fig. 4.   Original carrier image and scrambled image.

The original QR Code image which content is the text watermark information is shown in Fig. 1(a), the encryption result of it is shown in Fig. 1(b), both the size of the two images are 58×58.

After a great deal of experiments, the embedding positions we chosen are $\{9,10,11\}$ and the embedding strength $\lambda$ is 10. The positions they represents are in the Fig. 4, the watermarked image displays in Fig. 5. And the PSNR between the original and the watermarked image is 41.63, which guarantees the good imperceptibility [17].

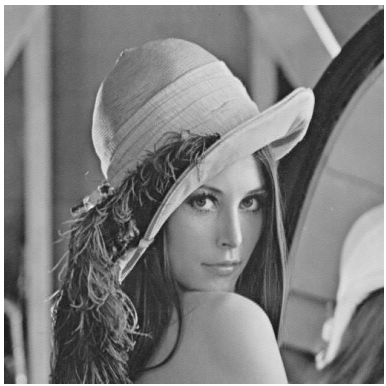

Fig. 5.   The watermarked image.

## C. Watermark attacks experiments

In order to test the robustness of our approach, we have used multiple attacks to the watermarked image such as JPEG compression (in different quality factor), Gaussian noise, image filtering, partially image cutting, image histogram equalization and so on. The PSNR described in Eq. (4) is also used to evaluate the degrees of the attacks in watermarked image. The attacked watermarked images are shown in Fig. 6(a-j).

The PSNR values of these attacked watermarked images compared with the original watermarked images, the QR Code images extracted from these attacked images, the corresponding NC values and the situations of QR code decoding are shown in Table 1. The QR Code decoding software is Psytec QR Code Editor V2.43.

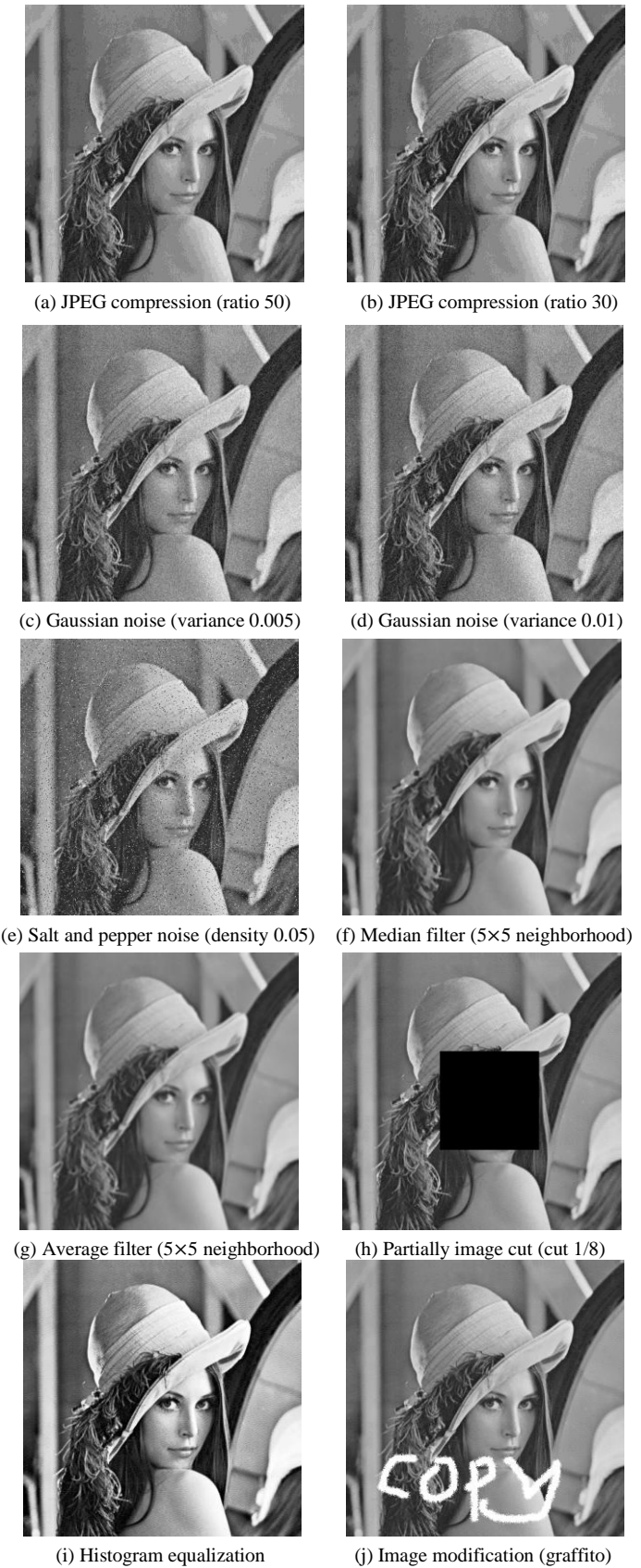

(a) JPEG compression (ratio 50)    (b) JPEG compression (ratio 30)

(c) Gaussian noise (variance 0.005)    (d) Gaussian noise (variance 0.01)

(e) Salt and pepper noise (density 0.05)    (f) Median filter (5×5 neighborhood)

(g) Average filter (5×5 neighborhood)    (h) Partially image cut (cut 1/8)

(i) Histogram equalization    (j) Image modification (graffito)

Fig. 6.   The attacked watermarked images.

TABLE I. THE RESULT OF WATERMARK EXTRACT

| Attacked Image | PSNR (dB) | Extracted QR Code | NC | Can be decoded? |
|---|---|---|---|---|
| Fig. 6(a) | 35.7832 | 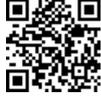 | 1 | Yes |
| Fig. 6(b) | 32.7764 | 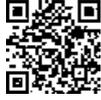 | 0.9769 | Yes |
| Fig. 6(c) | 23.0421 | 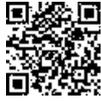 | 0.9585 | Yes |
| Fig. 6(d) | 20.0599 | 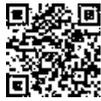 | 0.9157 | Yes |
| Fig. 6(e) | 18.3864 | 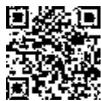 | 0.9391 | Yes |
| Fig. 6(f) | 11.6493 | 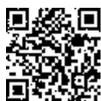 | 0.9718 | Yes |
| Fig. 6(g) | 11.7461 | 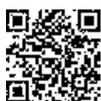 | 0.9685 | Yes |
| Fig. 6(h) | 14.7487 | 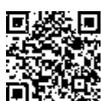 | 0.9441 | Yes |
| Fig. 6(i) | 42.0964 | 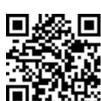 | 1 | Yes |
| Fig. 6(j) | 17.6441 | 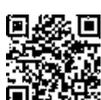 | 0.9535 | Yes |

From the above experimental results, it can be seen that this paper's watermarking approach has a high robustness for image lossy compression, noise pollution, image filtering, image enhancement and other common used image process methods.

Comparing with the general robust watermarking approaches which uses a meaningful binary image as the watermark image [18], the advantages of our proposes approach is that our watermark information can relatively contain longer text information and we can guarantee the content of the QR Code is the embedded text watermark information, as long as the extracted QR Code image can be decoded normally.

## V. CONCLUSION

The two-dimensional barcode technologies such as QR Code have been widely used in the whole world for its low cost, high reliability, high capacity, flexible encoding, etc. [9]. This paper, combines QR Code and digital watermarking technology together to propose a watermarking approach which has a strong robustness and security, and can obtain relatively more watermark information than other approaches. The simulation experiments prove these advantages. Thus, it has a certain application value on copyright protection and content authentication for digital media.